\documentstyle[11pt,newpasp,psfig]{article}
\begin{document}

\title{Dynamical Evolution: Spirals and Bars } 
\author{F. Combes}
\affil{DEMIRM, Observatoire de Paris, 61 Av. de l'Observatoire,
F-75 014, Paris, France}

\begin{abstract}
Non-axisymmetric modes like spirals and bars are the main
driver of the evolution of disks, in transferring angular
momentum, and allowing mass accretion. This evolution
proceeds through self-regulation and feedback mechanisms, 
such as bar destruction or weakening by a
central mass concentration, decoupling of a nuclear bar
taking over the gas radial flows and mass accretion, etc..
These internal mechanisms can also be triggered by interaction
with the environment. Recent problems are discussed,
like the influence of counter-rotation in the m=1 and 
m=2 patterns development and on mass accretion by a 
central AGN.
\end{abstract}


\section{ Bar formation and destruction}

\subsection{Rapid evolution of disks}

Galaxy disks are far from stationary. They are unstable, with time-scales 
that can be very short according to the radius. This ranges from
Myr in the centers to Gyr in the outer parts,
According to the environment, companions, mass accretion, etc.. 
normal modes can be excited. With the mass distribution of normal
nearby galaxies, the most rapidly growing mode is the m=2
(spirals and bars). Trailing waves transfer angular momentum to the outer parts, 
and concentrate the mass. This mass concentration itself changes
the development condition of the modes, and weakens them.
Alternatively, the disk is heated by waves and the perturbation fades 
away, until the next excitation.

\subsection{ Self-regulation}
\label{self-r}

The evolution of the disk is then controlled by recurrent waves,
through regulation and feedback processes. One of these
is related to gravitational instabilities, suppressed or
favored by heating and gas cooling. When the disk is cold
( Toomre Q parameter $<$ 1), and therefore unstable to spiral 
and bar waves, it develops non-axisymmetric perturbations and gravity 
torques, that transfer the angular momentum outwards.
The disk is progressively heated by the waves until 
the Q threshold is reached. A disk with only old stars 
cannot cool and will then remain stable, while a galaxy
rich in gas will continue to be unstable by gas cooling.
Young stars formed out of the gas reform a cool 
unstable stellar disk.

The gravity torques that drive the gas inwards play the role
of a viscosity; the galaxy disk can then be considered as an
accretion disk. An exponential stellar disk can then be built,
through star formation, if there is 
approximate agreement between the two time-scales, 
viscosity and star-formation (Lin \& Pringle 1987). This is
the case, since the two processes depend 
exactly on the same gravitational instabilities. 

The radial inflow of gas towards the center produces 
a mass accumulation, that can destroy the bar. This
occurs when about 5\% of the mass of the disk has sunk inside
the inner Lindblad resonance (Hasan \& Norman 1990, 
Pfenniger \& Norman 1990, Hasan et al 1993).
In the mean time, such a mass accretion in the center
has created a nuclear disk, which may decouple kinematically
from the rest of the disk; there can
develop either a secondary bar  (Friedli \& Martinet 1993)
or nuclear spirals (Barth et al. 1995, Regan \& Mulchaey 1999).
Alternatively, the decoupling may not occur, and the
central pattern rotates at the same velocity as the 
extended one (Englmaier \& Shlosman 2000). This
depends on the equivalent viscosity and star-formation
rates (Combes 1994, Sheth et al. 2000).
It is possible that the decoupling involves also a warp
(Schinnerer et al. 2000a,b).

After a bar has dissolved, the axisymmetric disk is devoid
of torques, and the new gas accreted stays in the extended disk.
If there is significant accretion with respect to the mass of
the disk, the latter becomes unstable and another bar forms.
 External interactions and mergers can intervene to destroy
the disk, but also to replenish it. The main agents to drive
matter to the center in galaxy interactions and mergers, 
are also the bars triggered by the tidal forces.

\subsection{Is the bar responsible for the fueling?}

There have been several observational works revealing
a correlation between nuclear activity and bars (Dahari 1984, 
Simkin et al 1980, Moles et al 1995). But the correlation is
weak and depends on the definitions of the samples,
the completion and other subtle effects. Near-infrared
images have often revealed bars in galaxies previously 
classified unbarred, certainly due to gas and dust effects.
However, Seyfert galaxies observed in NIR do not statistically
have more bars nor more interactions than a control sample 
(McLeod \& Rieke 1995, Mulchaey \& Regan 1997).

Peletier et al (1999) have recently re-visited the question,
and took a lot of care with their active and control samples.
Their Seyfert and control samples are different at 2.5 $\sigma$,
in the sense that Seyferts are more barred. They also measure
the bar strength by the observed axial ratio in the images.
In Seyferts, the fraction of strong bars is lower than in 
the control sample (Shlosman et al. 2000).
Although a surprising result a priori, this is not unexpected,
if bars are believed to be destroyed by central mass
concentrations (cf section \ref{self-r}).

Regan \& Mulchaey (1999) have studied 12
Seyfert galaxies with HST-NICMOS. Out of the 12, only 3 have 
nuclear bars but a majority show nuclear spirals. However their
criterium for nuclear bars is that there exist leading dust lanes
along this nuclear bar. This is not a required characteristic, since
these secondary bars in general are not expected to have ILRs
themselves.

\section{Lopsidedness, m=1 perturbations}

The $m=1$ perturbation is present in most galactic disks
(Richter \& Sancisi 1994), often superposed to the $m=2$ ones.
These perturbations can be of different nature and origin,
according to their scale (nuclear or extended disk),
or whether they involve the gaseous or stellar disks.  

Linear analysis, supported by numerical calculations,
show that gaseous disks rotating around a central mass,
are unstable for low value of the central 
mass (Heemskerk et al 1992). The instability disappears
when the central mass equals the disk mass.

Density waves for $m=2$ and higher generally use
the corotation amplifier, together with reflections at
resonances or boundaries, for a steady mode to grow. 
For $m=1$ in nuclear disk, the pattern speed is so low
that the corotation falls outside the disk. 
But, due to the special character of $m=1$ waves,
there exists another amplifier: the indirect term (Adams et al. 1989,
Shu et al. 1990). This is due to the off-centering of the center 
of mass. Long lasting oscillations of a massive nucleus
have been observed (Miller \& Smith 1992, Taga \& Iye 1998).

For extended disks, off-centered in an extended dark halo,
lopsidedness could survive, if the disk remains in the region
of constant density (or constant  $\Omega$(r)) of the 
halo (Levine \& Sparke 1998).

In the special case of a stellar nuclear disk around
a massive black-hole, it is possible that self-gravity
is sufficient to compensate for the differential precession
of the nearly keplerian orbits, and that a long-lasting
$m=1$ mode develops, or is maintained long after an
external excitation (Bacon et al. 2000). This could
be the explanation of the double nucleus observed 
for a long time in M31 (Bacon et al. 1994, Kormendy \& 
Bender 1999), and for which an eccentric disk model
has been proposed (Tremaine 1995, Statler 1999).

\begin{figure}
\centerline{\psfig{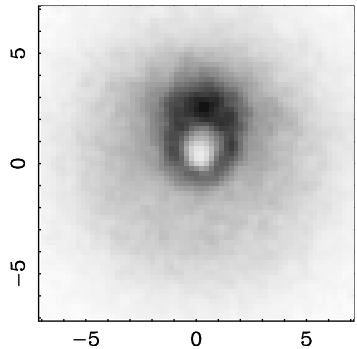}}
\caption{Face-on surface density of the nuclear stellar disk of M31, at epoch
28.8 Myr of the simulation. The scale is in pc (from B00). }
\label{cont_m19fpc}
\end{figure}

\begin{figure}
\centerline{\psfig{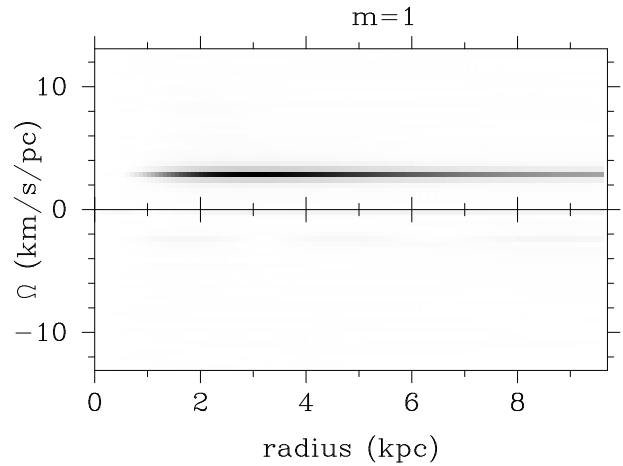}}
\caption{Pattern speed as a function of radius, in units of km/s/pc, for
the $m=1$ mode, between 28.8 and 43.2 Myr of the M31 simulation.
The pattern speed slightly slows down with time (from B00).}
\label{pow_m19fpc}
\end{figure}

\begin{figure}
\centerline{\psfig{figure=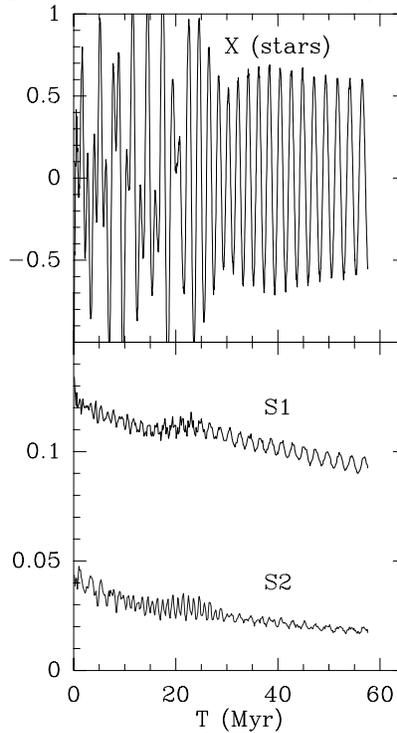,width=5cm,bbllx=32mm,bblly=15mm,bburx=185mm,bbury=10cm,angle=-90}}
\caption{{\bf top:} Coordinates (X: full line) of the center of gravity of
the stellar disk, as a function of time. 
{\bf bottom:} Intensity of the $m=1$ (S1, solid line) and
$m=2$ (S2, light line) Fourier components of
the potential (more exactly the corresp. components of the tangential force normalised by
the radial force), for the best fit model of M31 (from B00).}
\label{p1cet_m19fpc}
\end{figure}

In this $m=1$ mode, the maximum density is obtained at
the apocenter of the aligned elongated orbits (see fig \ref{cont_m19fpc}).
The pattern speed, equal to the orbital frequency of the barycentre
of the stellar disk, is slow (3km/s/pc, fig \ref{pow_m19fpc}), 
with respect to the orbital frequency of the stars themselves 
(250km/s/pc in the middle of the disk).
The excitation of the $m=1$ perturbation can then last
more than  3000 rotation periods (fig \ref{p1cet_m19fpc}).

\section{Counter-Rotating Components}

The phenomenon of counter-rotating components is
a tracer of galaxy interactions, mass accretion or mergers.
It has been first discovered in ellipticals with kinematically 
decoupled cores (likely to be merger remnants, 
e.g. Barnes \& Hernquist 1992).
It has been observed also in many spirals; either two
components of stars are counter-rotating, or the gas with respect
to the stars, or even two components of gas,
in different regions of the galaxies
(Galletta 1987, Bertola et al 1992). In the
special case of NGC 4550 (Rubin et al. 1992), two
almost identical counter-rotating stellar disks are superposed 
along the line of sight.

These systems pose a number of questions, first from their
formation scenario, but also about their stability, their
life-time, etc.. Do special waves and instabilities 
develop in counter-rotating disks, and 
does this favor central gas accretion?

\subsection{Stability}

First, it appears that the counter-rotation (CR) can bring more stability.
Even a small fraction of CR stars has a stabilising influence with 
respect to bar formation ($m=2$), since the disk has then 
more velocity dispersion (Kalnajs 1977).
But a one-arm instability is triggered, for a comparable 
quantities of CR and normal stars. This comes from the 
two-stream instability in flat disks,
similar to that in CR plasmas (Lovelace et al. 1997). There
develop two $m=1$ modes in the two components,
with energies of opposite signs: the negative-E mode 
can grow by feeding energy in the positive-E mode.

A quasi-stationary one-arm structure forms, and lasts 
for about 1 to 5 periods 
(Comins et al. 1997). The structure is first leading, 
than trailing, and disappears.
The formation of massive CR disks in spirals
has been studied by Thakar \& Ryden (1996, 1998).

\subsection{CR with gas}

The presence of two streams of gas in the same plane 
will be very transient: strong shocks will occur, producing
heating and rapid dissipation. The  
gas is then driven quickly to the center. But the two streams of gas 
could be in inclined planes, or at different radii.
This is the case in polar rings, discovered in
0.5 \% of all galaxies (Whitmore et al. 1990).
After correction for selection effects (non optimal viewing, 
dimming, etc..) 5\% of all S0 would have polar rings.

\begin{figure}
\centerline{\psfig{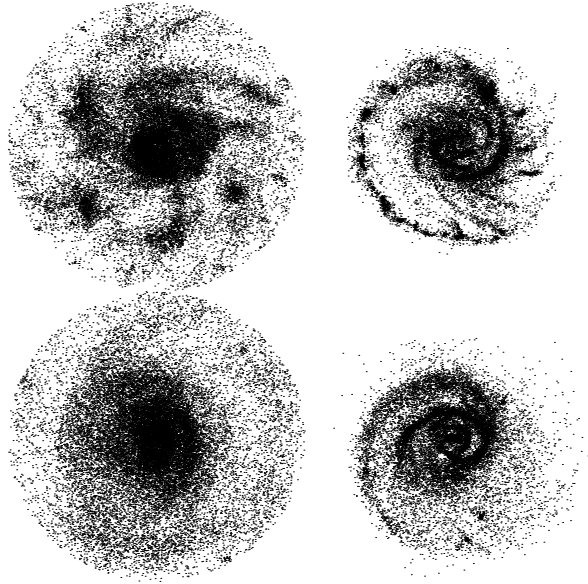}}
\caption{Particle plots of the stars (left) and the gas (right)
for the NGC 3593 N-body simulations,
at successive epochs 200 and 400 Myr.
Particles are plotted until a radius of 6.25 kpc.
The majority of stars rotate in the direct sense, while the gas is
retrograde (from G00).} \label{cin_ndm}
\end{figure}

\begin{figure}
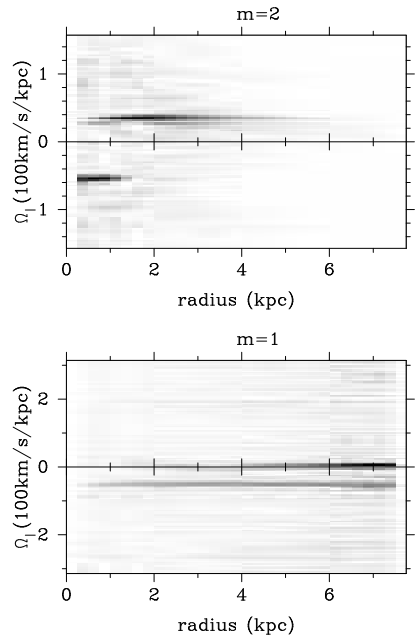

\centerline{\psfig{figure=combesf_f5a.ps,width=53mm,bbllx=15mm,bblly=1cm,bburx=105mm,bbury=12cm,angle=-90}}
\centerline{\psfig{figure=combesf_f5b.ps,width=53mm,bbllx=15mm,bblly=1cm,bburx=105mm,bbury=12cm,angle=-90}}
\caption{Pattern speed as a function of radius, in units of 100km/s/kpc, for
the $m=2$ mode ({\bf top}), and the  $m=1$ mode   ({\bf bottom}),
for the NGC 3593 simulation (from G00).}
\label{pow_ndm}
\end{figure}

If there is only one gas stream, the problem is more similar
to the two-stream instabilities of stellar disks mentioned above.
However the gas is cooling, and is not easy to stabilise
against $m=2$ components. Both $m=1$ and $m=2$ may be
present simultaneously in these systems. This is the case of
the galaxy NGC 3593, composed of two CR stellar disks
(Bertola et al 1996): in an extended stellar disk, is 
embedded a CR nuclear disk, possessing co-rotating gas.
 The molecular component associated with this nuclear
disk reveals both a nuclear ring and a one-arm spiral
structure outside of the ring (Garcia-Burillo et al. 2000).
N-body simulations have shown that both structures can
be explained by the superposition of  $m=1$ and $m=2$ 
in the gas component, the ring being formed at the ILR
of the bar (fig. \ref{cin_ndm}). In the $m=2$ pattern, 
two counter-rotating bars develop (fig \ref{pow_ndm}).

\section{ Conclusions}

Galactic disks are not stationary, they evolve greatly on  
time-scales of the order of a few dynamical times. In normal
disks, there is a preponderance of $m=2$ spirals and bars.
Spirals are the most transient features, but bars also 
come and go, and several bar episodes could occur in a Hubble time.

According to the mass concentration in the disk, 
a secondary bar can decouple and rotate with
a higher pattern speed. This is important for the
ultimate fueling of gas to the nucleus. Observational
evidence of correlation between nuclear activity and bars
is still controversial. It might be a question of spatial resolution,
or the fact that bars are destroyed by massive central black holes.

Lopsidedness is also wide-spread in galactic disks, but
much less studied. The nature and origin of the $m=1$ waves
are diverse, either particular to a purely gaseous disk,
or a nuclear stellar disk. They could be slowly damped modes 
(excited by companions), or two-stream instabilities in
counter-rotating systems. There is also the special case
of rapid off-centring instabilities (oscillations of the center
of mass), or nearly keplerian eccentric disks near a 
central massive black hole.

Non-axisymmetric perturbations are transferring angular
momentum outwards, and result in rapid evolution of galactic 
disks. It is likely that continuous disk reformation occurs
all over the Hubble time, through mass accretion.

\acknowledgements {I thank the organisers for a splendid conference,
and my collaborators for letting me use common work prior to
publication.}


\end{document}